\documentclass[twocolumn, floatfix]{aastex631}
\usepackage{lipsum}

\usepackage{amsmath}
\usepackage{graphicx}

\usepackage{float}
\usepackage{natbib}
\usepackage{soul}
\usepackage{array}
\usepackage{booktabs}

\begin{document}

\title{Direction-dependent effects on global 21-cm detection}

\author{Yash Agrawal}
\affiliation{Raman Research Institute, CV Raman Avenue, Sadashivanagar, Bangalore, 560080, India}

\author{K. Kavitha}
\affiliation{Raman Research Institute, CV Raman Avenue, Sadashivanagar, Bangalore, 560080, India}

\author{Saurabh Singh}
\affiliation{Raman Research Institute, CV Raman Avenue, Sadashivanagar, Bangalore, 560080, India}


\begin{abstract}

Cosmic dawn represents critical juncture in cosmic history when the first population of stars emerged. The astrophysical processes that govern this transformation need to be better understood. The detection of redshifted 21-cm radiation emitted from neutral hydrogen during this era offers a direct window into the thermal and ionization state of the universe. This emission manifests as differential brightness between spin temperature and the cosmic microwave background (CMB). SARAS experiment aims to detect the sky-averaged signal in the frequency range 40-200~MHz. SARAS's unique design and operation strategy to float the antenna over a water body minimizes spectral features that may arise due to stratified ground beneath the antenna. However, the antenna environment can be prone to configuration changes due to variations in critical design parameters such as conductivity and antenna tilts. In this paper, we connect the variations in antenna properties to signal detection prospects. By using realistic simulations of a direction and frequency-dependent radiation pattern of the SARAS antenna and its transfer function, we establish critical parameters and estimate bias in the detectability of different models of the global 21-cm signal. We find a correlation between the nature of chromaticity in antenna properties and the bias in the recovered spectral profiles of 21-cm signals. We also find stringent requirements for transfer function corrections, which can otherwise make detection prospects prohibitive. We finally explore a range of critical parameters that allow robust signal detection.

\end{abstract}


\section{Introduction} \label{sec: intro}

One of the poorly understood aspects of cosmic evolution is the formation of the first stars and galaxies. Cosmic dawn and subsequent epoch of reionization (EoR) were the periods in cosmic history when the first population of stars came into existence, marking the end of the dark ages \citep{2001PhR...349..125B,2003Sci...300.1904M}. Ionizing radiation from these sources eventually depleted the neutral hydrogen in the universe \citep{2013ASSL..396...45Z}. In the $\Lambda$CDM model, the thermal history of baryons during the dark ages is well-constrained and primarily governed by the universe's expansion, proton and electron recombination, and interaction of the residual free electrons and cosmic microwave background (CMB). However, when the first sources of radiation formed, they changed the trajectory of the thermal and ionization state of the intergalactic medium (IGM) \citep{2005SSRv..116..625C}. The astrophysical processes during cosmic dawn and EoR need to be better understood \citep{2006ARA&A..44..415F}.

One promising technique for studying the thermal and ionization properties of the universe during this era is the 21-cm spin-flip line from neutral hydrogen \citep{1999A&A...345..380S,2006PhR...433..181F,2012RPPh...75h6901P,2004MNRAS.347..187F}. Beyond cosmic dawn, this spin transition firmly links itself with the thermal temperature of hydrogen gas under the assumption of strong coupling due to the Wouthuysen Field (WF) effect \citep{1952AJ.....57R..31W,1959ApJ...129..536F}. Detection of sky-averaged 21-cm radiation during this era can place stringent constraints on several astrophysical parameters \citep{2008PhRvD..78j3511P,2010PhRvD..82b3006P,2003ApJ...596....1C}. This radiation, observed as the global 21-cm signal,  presents itself as a mean departure of spin temperature from CMB temperature at the rest frame frequency of 1420 MHz. However, the sky averaged 21-cm signal component corresponding to cosmological redshift $6 < z < 35$ is stretched to metre wavelength and lies between $\sim$ 40 MHz to 200 MHz. Several ground-based experiments such as LEDA~\citep{2015ApJ...799...90B}, EDGES \citep{2017ApJ...847...64M}, SARAS~3 \citep{2021arXiv210401756N}, SCI-HI~\citep{2014ApJ...782L...9V}, REACH \citep{2022NatAs...6..984D}, $PRI^{Z}M$ \citep{2019JAI.....850004P}, and MIST \citep{2023arXiv230902996M} are attempting to detect the cosmological signal. Notably, EDGES collaboration claimed a tentative detection of a flattened Gaussian profile \citep{bowman2018absorption}, which has been refuted by SARAS~3 observations \citep{2022NatAs...6..607S}. In view of ground-based challenges, several space-based experiments are also proposed, including PRATUSH \citep{2023ExA...tmp...49S},  DAPPER \citep{2021RSPTA.37990564B}, DSL \citep{2022ApJ...929...32S}, and LuSEE `Night' \citep{2023arXiv230110345B}.

The SARAS experiment is custom-designed to observe the 21-cm signal from cosmic dawn and EoR. In its current third edition, SARAS is optimized to operate in the frequency range of 40~MHz to 90~MHz. SARAS~3 antenna sits on a styrofoam raft floating on a water body. The motivation behind the choice of water as a medium directly below the antenna to a considerable depth is that water provides a homogeneous medium, avoiding significant spectral features. Additionally, placing an antenna over water with relative permittivity of ~80, rather than over the dry ground with relative permittivity of ~4-5, would achieve a higher radiation efficiency because of the higher permittivity of water relative to the soil \citep{2021ITAP...69.6209R}. However, this configuration makes the antenna susceptible to several perturbations like variations in raft height, tilts, and water's conductivity. The systematics involved will be affected by both the orientation of the antenna and variations in water properties.

The primary goal of this work is to assess the capability of the SARAS~3 radiometer in constraining different modes of global 21-cm signals under these perturbations. A direct approach to achieving this involves simulating realistic sky observations and attempting to recover signals of different spectral profiles.

Section \ref{sec: antennas} we detail the electromagnetic simulations carried out for the SARAS~3 antenna. This will be followed by Section \ref{sec: sky smooth}, where we investigate the spectral nature of the sky model that we are using to simulate our data using an achromatic beam. In Section \ref{sec: beam smooth}, we simulate data under different perturbations of SARAS~3 beam and assess the systematics introduced due to beam chromaticity. We further correlate results with primary beam derivatives .We further evaluate the systematics caused by a miscalibrated return loss of the antenna in Section \ref{sec: S11 correction}. Finally, in Section \ref{sec: signal extraction}, we inject a range of 21-cm signals and employ data models and assess the signal extraction capabilities of the current version of SARAS under different perturbations. Finally conclusions, derived from the cumulative outcomes of all previous sections, are presented in Section \ref{sec: conclusion}. These interpretations, while primarily centered around SARAS~3 simulation cases, hold general applicability across experiments of similar nature.

\section{Electromagnetic simulations of the antenna} \label{sec: antennas}
\subsection{Description of the SARAS~3 antenna}
The SARAS~3 antenna \citep{2021ITAP...69.6209R} floats on a freshwater lake medium below, compared to previous deployments over soil medium. Such media avoid systematic effects due to multipath propagation between different layers and make it easier to model the antenna and its environments electromagnetically. The medium is characterized by its dimensions, conductivity, and dielectric constant. This is in contrast to stratified ground layers, where a detailed profile of each layer is required for large volumes \citep{2022MNRAS.515.1580S, 2023arXiv231007741M}. The baseline design of SARAS~3 was decided to maximize its spectral smoothness in coupling sky power to the receiver, governed by the return loss. Further, the design minimized variation of its primary beam with frequency as chromatic beams can introduce significant spectral structures complicating foreground modeling \citep{2023MNRAS.521.3273S, 2021AJ....162...38M}.

\begin{figure}[!h]
    \centering
    \includegraphics[width=1\columnwidth]{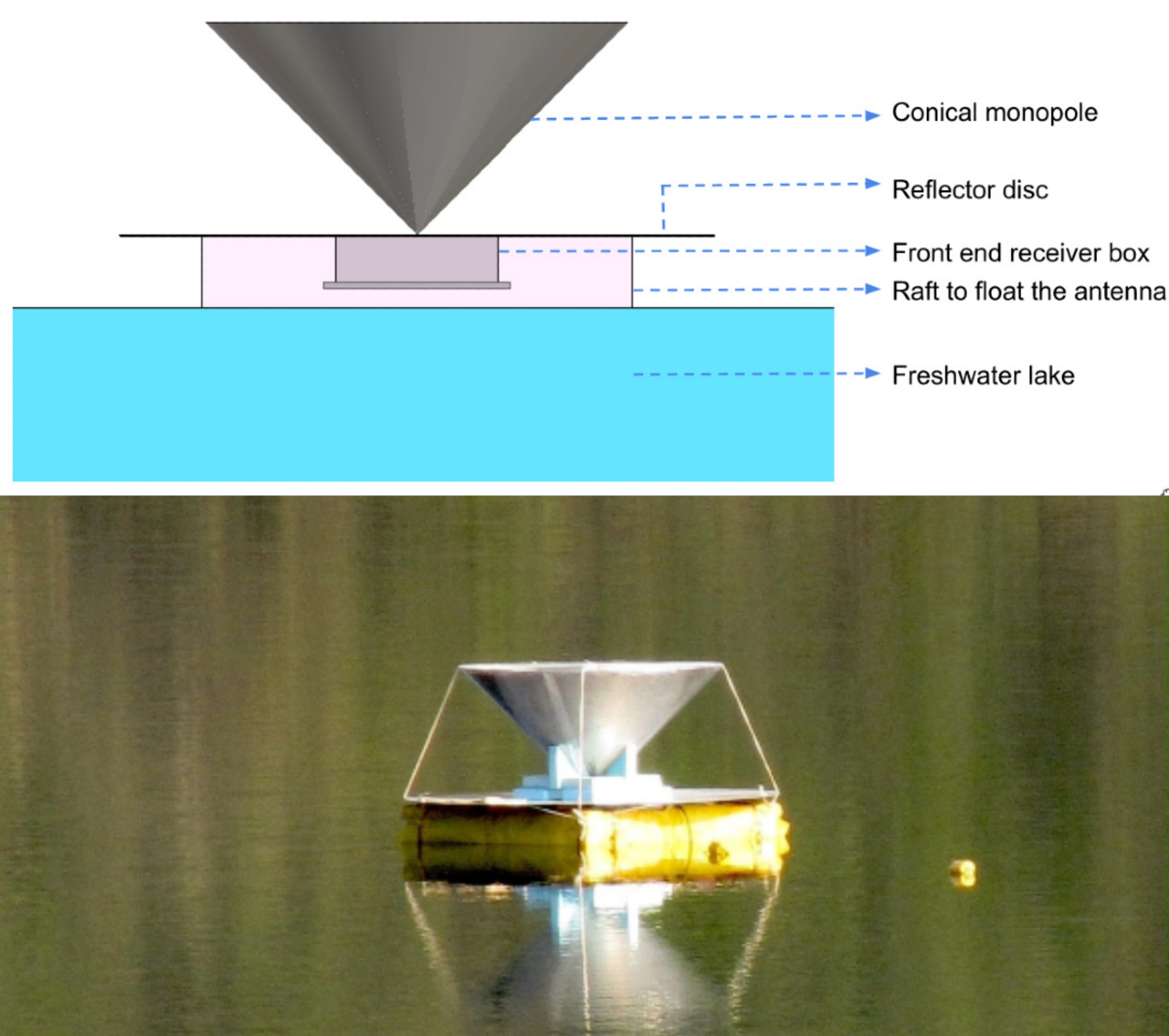}
    \caption{SARAS~3 antenna floating on ideal lake environment}
    \label{fig: Ideal_SARAS3_antenna}
\end{figure}

However, changes in the environment surrounding the antenna and perturbations in the medium on which the antenna is deployed can introduce systematic errors. We find that antenna properties are most susceptible to the conductivity of the water medium, the height of the antenna above the surface of the water, and the orientation/tilt of the antenna with respect to the horizon. Figure \ref{fig: Ideal_SARAS3_antenna} shows the schematic of the antenna floating over the water medium.

Electromagnetic simulations are performed using CST \footnote{Computer Simulation Technology— \href{https://www.cst.com}{www.cst.com}} microwave to simulate these varying environments and capture their effects. CST's Integral equation solver and Multilayer solver can be used for these simulations. CST's multilayer solver was preferred for analysis as it was found to be more efficient in performing the simulation compared to the integral equation solver in terms of computation time. It also allowed for modeling multiple layers, including the lake and the lake bed layer, into simulations when required.

\subsection{Baseline antenna configuration and its variations}
 The antenna was initially designed for deploying over freshwater medium of low conductivity $< 0.05~S/m$ using an electromagnetically transparent raft whose height is 200 mm \citep{2021ITAP...69.6209R}. However, the conductivity of the water medium may vary according to the location, as SARAS~3 can be deployed at different sites depending on their radio quietness. Also, the water medium over which the antenna is deployed might not be stable during windy weather conditions. An unstable water surface might lead to vertical movement of the raft on the surface of the water, in addition to introducing variations in the orientation of the antenna axis, tilting it in different directions away from the zenith. As discussed above, we assess the effects of changing the following three parameters: water conductivity, antenna reflector height from water, and the tilt angle of the raft, which consequently affects the antenna's tilt. Based on realistic field conditions, we vary conductivity from that of fresh water to saline, in the range of $0.001$ to $4~S/m$. The height of the antenna above water is varied from 200~mm to 50~mm. Finally, the tilt of the antenna is changed from 0 to 6 degrees. The conductivity of $0.001~S/m$ with an antenna 200~mm above water with no tilt is taken as the baseline configuration. We simulate antenna characteristics, namely its primary beam pattern and return loss, over a range of the above-mentioned degrees of variations. We evaluate the primary beam over 40-90~MHz, spaced every 1~MHz, with open add space boundary conditions followed by a perfectly matched layer. The Perfectly Matched Layer is placed at a distance of $\lambda/4$ away from the antenna where wavelength $\lambda$ is taken at the lowest frequency of the simulation band.

\section{Investigating the Smoothness of the Sky Model} \label{sec: sky smooth}

\subsection{Sky Model: GMOSS} \label{subsec: GMOSS}
Before investigating the impact of the antenna beam variation on the observed sky, we begin with modeling the intrinsic foregrounds weighted by an achromatic beam. This initial assessment with an achromatic beam helps in understanding the intrinsic spectra of the sky model before any systematics are introduced by actual antenna beams.
We have used a physically motivated sky model to simulate the data. Global MOdel for the radio Sky Spectrum (GMOSS) predicts radio brightness across the frequency range of 22~MHz to 23~GHz by constructing a physically motivated model of the relevant radiative processes \citep{rao2016gmoss}.
These processes involve galactic and extra-galactic emissions, which are constrained by all-sky measurements. Previous works have shown that these foregrounds can be modeled with a maximally smooth function \citep{2017ApJ...840...33S}, unlike the global signal, which is predicted to have spectral features and inflection points. Here, we generate mock observations using GMOSS and estimate levels of spectral features intrinsic to the foregrounds in 40-90~MHz. 

\subsection{Data Simulation and analysis}
To investigate the spectral smoothness of the sky model GMOSS, we generate beam-weighted sky temperature from the sky map for a primary beam of form $cos^{2}\theta$. This analysis covers a frequency range of 40 to 90~MHz, where $\theta$ is the elevation angle. Such achromatic forms of primary beams can be realized using electrically small antennas \citep{2021ITAP...69.6209R, 2005atad.book.....B}. The achromatic nature of the $\cos^{2}$ beam preserves the spectral integrity of the intrinsic sky spectrum. The beam weighted sky is given by,

\begin{equation}\label{eq: simulated_data_cosine}
    T_{w}(\nu,t) = \frac{\int_{\phi = 0}^{\phi = 2\pi}\int_{\theta = 0}^{\theta = \pi/2} T_{F}(\theta, \phi, \nu, t) G(\theta, \phi, \nu) \;\ \sin\theta d\theta  d\phi}{\int_{\phi = 0}^{\phi = 2\pi}\int_{\theta = 0}^{\theta = \pi/2} G(\theta, \phi, \nu) \;\ \sin\theta d\theta  d\phi}
\end{equation}

where, $T_{F}(\theta, \phi, \nu, t)$ is brightness of the foregrounds from GMOSS and $G(\theta, \phi, \nu) =cos^{2}\theta$.

We generate a beam-weighted sky over a 24-hour period at a location in the Hanle region of India (32°48'46"N 78°52'16" E). Figure \ref{fig: cosine_sky_residuals} shows the GMOSS sky (panel (a)) and beam-weighted foreground variation over 24 hours (panel (b)). Note that we do not consider solar or ionospheric effects since our primary aim is to evaluate foreground mode mixing.
Subsequently, we calculate the average brightness along the temporal axis as shown in Figure \ref{fig: cosine_sky_residuals} panel (c). We further introduce a subdominant thermal noise of sub-milli kelvin root mean square (r.m.s).  Therefore, such a setup provides robust estimates of chromaticity in intrinsic foregrounds. In this and the upcoming sections, we will utilize this approach to produce simulated data for all variations of antenna configurations.

We use a logarithmic polynomial model for the antenna temperature \citep{2010PhRvD..82b3006P,2022NatAs...6..607S,bowman2018absorption} as given by,

\begin{equation} \label{eq: log_log_poly}
    T_{w}(\nu) =  10^{\sum_{n=0}^{m} \alpha_{n} (\log_{10} \nu)^n} ,
\end{equation}

where $T_w(\nu)$ is beam-weighted foreground temperature, $\alpha_{n}$ is the coefficient associated with each term in the polynomial in the log domain with $(m+1)$ terms.
The polynomial in this fit is constrained to be maximally smooth. A maximally smooth polynomial is a class of constrained functions that do not have inflection points or zero crossings in higher-order derivatives, i.e.,

\begin{equation}\label{eq: MS_condition}
    \frac{d^{m}y}{dx^{m}} \geq 0 \quad \text{or} \quad \frac{d^{m}y}{dx^{m}} \leq 0 ,
\end{equation}

where $m$ denotes $m^{th}$ derivative of the function $y$ for $m>2$.

 The same formalism has been used in the work of \citet{2017ApJ...840...33S} and further refined in \cite{2021MNRAS.502.4405B}
 
 Basinhopping \citep{doi:10.1021/jp970984n} is employed to optimize the parameters, minimizing the variance as shown in Equation \ref{eq:chi_sq}. The optimization process involves fitting polynomials of various orders. The requirement for maximal smoothness restricts the fit from exhibiting any complex spectral structures. This ensures that going to a higher order of the polynomial does not fit out the 21-cm signal, as is often the case with unconstrained polynomials. We decide the order by recording the reduction in the residual r.m.s with each additional term and stopping when the model accuracy hits the thermal noise floor or when the r.m.s saturates due to the constrained nature of the polynomial. The residuals thus obtained give an estimation of the spectral smoothness of the sky model. The residuals' r.m.s value is used as an indicator to assess the smoothness of fit.  

\begin{equation}\label{eq:chi_sq}
    \sigma^2 = \frac{\sum_{i=1}^{N}\left(10^{\sum_{n=0}^{m} \alpha_{n} (\log_{10} \nu)^n} - T_{w}(\nu)\right)}{N}
\end{equation}

where N is the number of data points.

\subsection{Results}
\begin{figure*}
    \centering
    
    \includegraphics[scale=0.5]{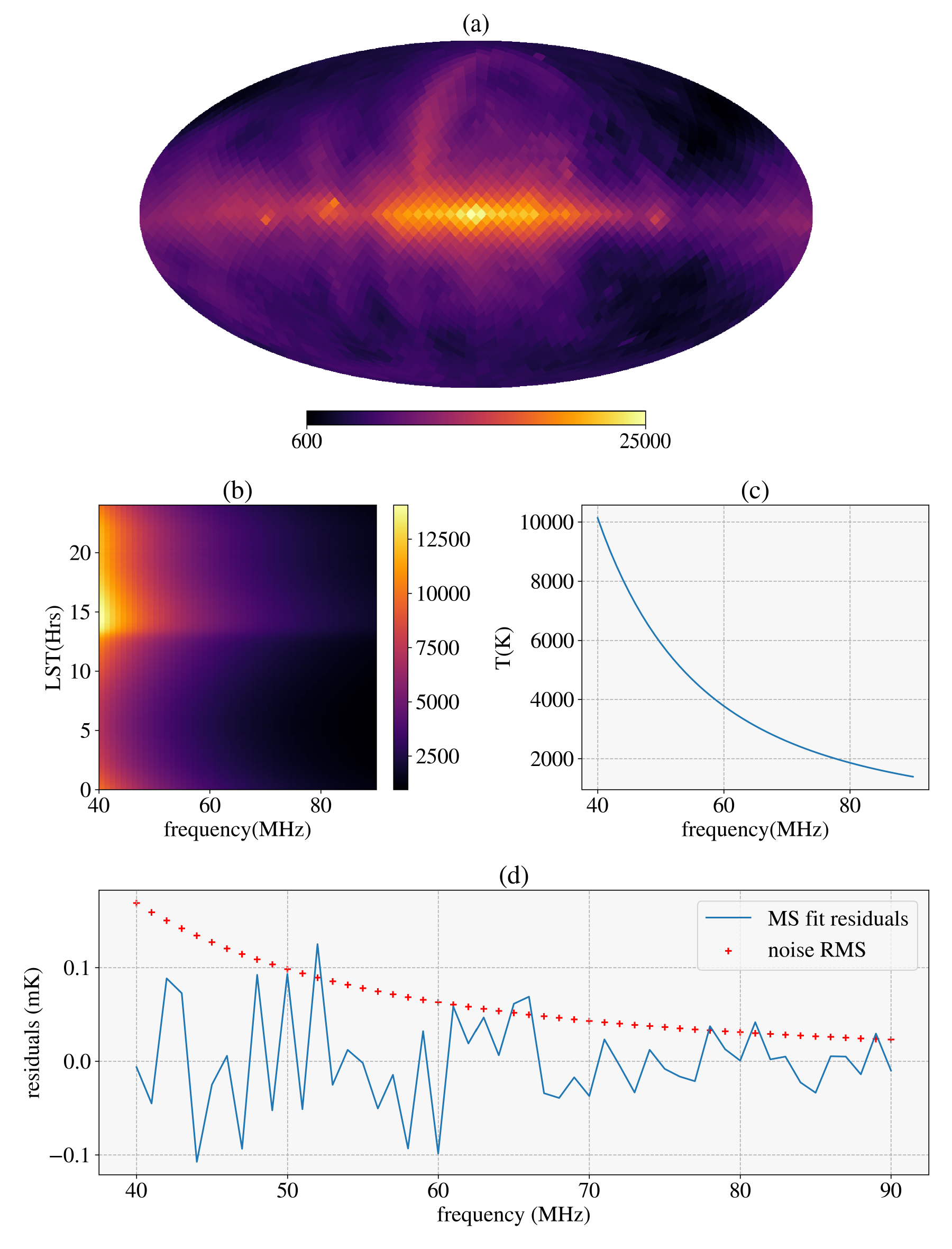}
    \caption{(a) GMOSS sky map Mollweide projection at 80~MHz, (b) Variation of beam-weighted sky with time and frequency, for the primary beam of form $\cos^2\theta$. Color represents brightness temperature in kelvin units for both panels (a) and (b); (c) Time-averaged foregrounds (d) Residuals (blue) after fitting maximally smooth function to panel (c), red plus-markers show the noise r.m.s.}
    \label{fig: cosine_sky_residuals}
\end{figure*}

We find that the residuals, after subtracting the fit, go down to the thermal noise level. Figure \ref{fig: cosine_sky_residuals}, panel~(c) shows the time-averaged foregrounds, while panel (d) shows the residuals obtained after fitting the simulated data with the model in Equation \ref{eq: log_log_poly}. Therefore, with a wide achromatic beam, foregrounds are demonstrated to be spectrally smooth to at least sub-mK level. Therefore, foregrounds, if observed through achromatic beams, can allow separation of the global 21-cm signal. This further implies that since achromatic beam does not add any spectral features to the inherent sky spectra, any excess spectral features through realistic beams in the following sections can be attributed to the chromaticity in the primary beam.

\section{Foregrounds as seen through realistic beams} \label{sec: beam smooth}
In Section \ref{sec: sky smooth}, we demonstrated the spectral smoothness of GMOSS. We also infer, from the magnitudes of residuals, that the faint global signal can, in principle, be separated from the bright foregrounds. In this section, we replicate the procedure outlined in Section \ref{sec: sky smooth}, but with the convolution of the sky model using various perturbations applied to the SARAS~3 beam, capturing realistic variations.

\subsection{Baseline vs. Perturbed Beam}
\begin{figure*}
    \centering
    \includegraphics[scale=0.4]{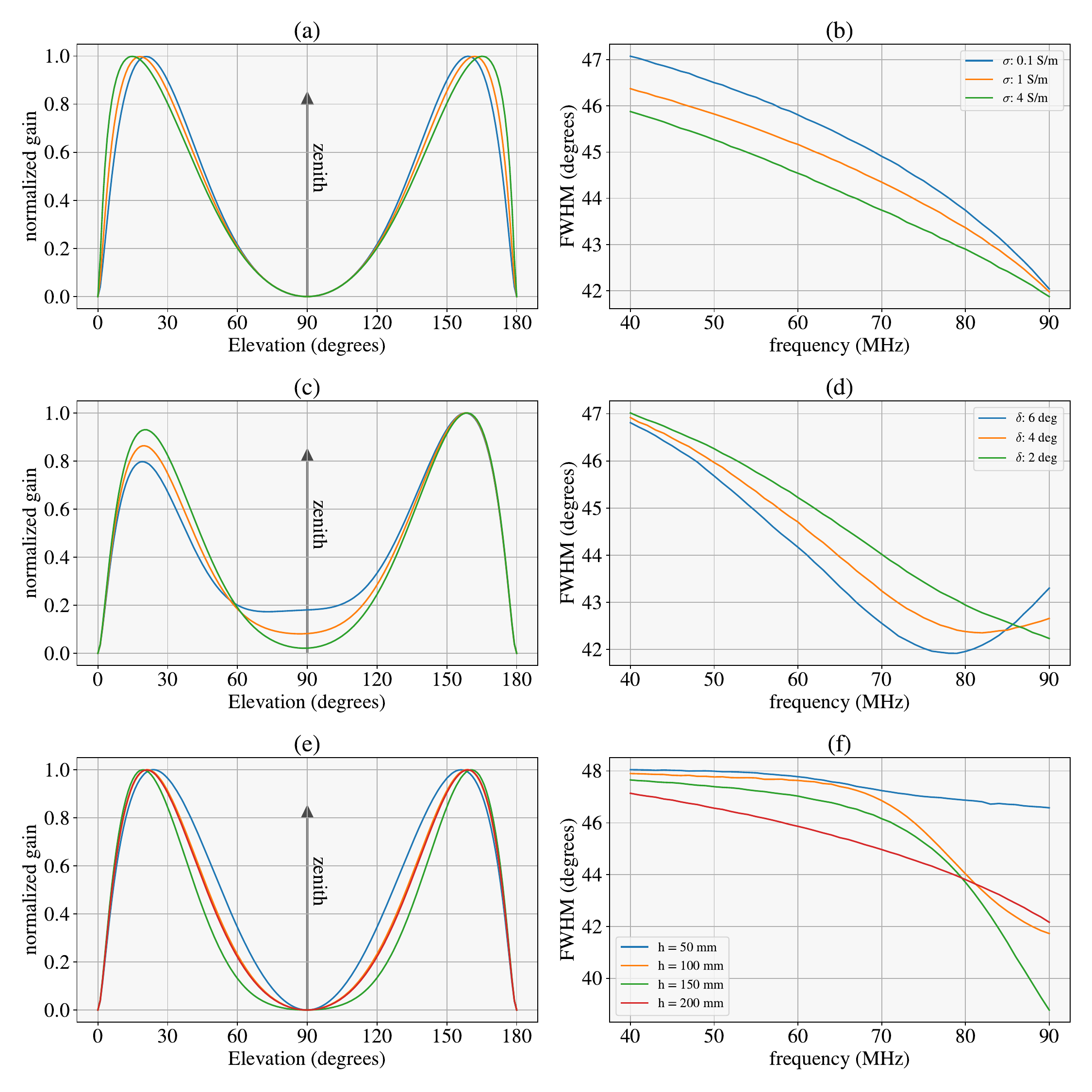}
    \caption{Panels (a), (c), and (e) show the spatial variation of the beams with changes in conductivity, tilt, and raft height at 70~MHz, respectively. Panels (b), (d), and (f) show the beam full-width-at-half-maximum (FWHM) over the frequency range 40~MHz to 90~MHz.}
    \label{fig: pole_plot_beam}
\end{figure*}

As mentioned in Section \ref{sec: antennas}, the SARAS~3 antenna sits on the raft floating above a deep water body under ideal conditions. SARAS~3 is optimized for low-conductivity water. The beam properties depend on the antenna's orientation and the surrounding medium's electrical properties. However, several perturbations could change the nature of the beam and other antenna characteristics. These variations include changes in the conductivity, height of the raft and orientation of the antenna. Our objective is to examine the systematic effects introduced by these perturbations. Table \ref{tab: residuals} summarizes the different cases simulated for the different variations of the parameters. In Figure \ref{fig: pole_plot_beam}, we show the spectral and spatial changes in the beam under different conditions.  We note that the beam width varies with frequency and perturbations, while the azimuthal symmetry of the primary beam is preserved in all the cases except the tilt variation.

\begin{table*}
    \centering
    \begin{tabular}{|c|c|c|c||c|c|}
    \hline
    Case No & Conductivity (S/m) & Tilt (°) & Raft Height (mm) & Residual r.m.s (mK) & Residuals with miscalibrated S11 (mK)  \\
    \hline\hline
    1 & 0.001 & 0 & 200 & 0.05 & 0.05\\
    \hline
    2 & 0.002 & 0 & 200 & 0.05 & 0.06\\
    \hline
    3 & 0.005 & 0 & 200 & 0.04 & 2\\
    \hline
    4 & 0.01 & 0 & 200 & 0.09 & 11\\
    \hline
    5 & 0.05 & 0 & 200 & 0.03 & 3\\
    \hline
    6 & 0.1 & 0 & 200 & 0.04 & 8\\
    \hline
    7 & 1 & 0 & 200 & 0.07 & 171\\
    \hline
    8 & 4 & 0 & 200 & 0.03 & 7\\
    \hline
    9 & 0.001 & 2 & 200 & 0.05 & 154\\
    \hline
    10 & 0.001 & 4 & 200& 0.05 & 119\\
    \hline
    11 & 0.001 & 6 & 200 & 0.09 & 65.7\\
    \hline
    12 & 0.001 & 0 & 150 & 51.5 & 294\\
    \hline
    13 & 0.001 & 0 & 100 & 56.14 & 3912\\
    \hline
    14 & 0.001 & 0 & 50 & 0.06 & 8076\\
    \hline \hline
    \end{tabular}
    \caption{Different cases for generating perturbed SARAS primary beams. Columns 2 -- 4 are the parameters used in the simulation, namely, water conductivity, antenna tilt, and raft height. The same case numbers will be used throughout the paper to refer to specific SARAS beam cases. Column 5 lists the residuals obtained after modeling foregrounds as described in Section \ref{sec: beam smooth}. Column 6 shows the residuals obtained after fitting foregrounds that have miscalibrated S11, as described in Section \ref{sec: S11 correction}}.
    \label{tab: residuals}
\end{table*}

\subsection{Data Simulation and Analysis}
To simulate the data, following Equation \ref{eq: simulated_data_cosine}, we take $T_{F}(\theta, \phi, \nu)$ to be the brightness of the foregrounds from GMOSS, however $G(\theta, \phi, \nu)$ is replaced with the realistic SARAS beam cases as listed in Table \ref{tab: residuals}.

We introduce a free constant parameter, c, in our model for the beam-weighted sky generated via SARAS beam perturbations. The antenna temperature for the SARAS beam cases is given by

\begin{equation}\label{eq: log_log_poly_DC}
T_{w}(\nu) =  10^{\sum_{n=0}^{m} \alpha_{n} (\log_{10} \nu)^n} + c ,
\end{equation}

where $T_{w}(\nu)$ is antenna temperature for the beam-weighted sky with SARAS beam, which is being modeled as a logarithmic polynomial and a constant parameter $c$. The logarithm of Bayesian evidence for the model with and without the free constant parameter differs by 20 for baseline SARAS case (1) as listed in Table \ref{tab: residuals}, providing compelling support for the constant parameter as the more suitable choice for the data \citep{doi:10.1080/01621459.1995.10476572} with realistic beams.

\subsection{Results} \label{subsection : purterbed beam results}
The r.m.s of residuals presented in Table \ref{tab: residuals} reveal that mostly all beam scenarios reach the noise floor level. In addition to averaging all 24 hours, we also explored the possibility of averaging for 8 and 12 hour LST-bins. Naturally, we see elevated residuals for the LST-bins where Galaxy is dominant. However, r.m.s. of residuals for baseline SARAS case (1) continue to be sub-mK in all the bins. The results presented herein correspond to 24-hour average. Nevertheless, unlike $cos^2\theta$ beam, these instances necessitate one inflection point that indicates excess spectral complexity. This could potentially arise due to systematics induced by beam chromaticity. We find that the baseline SARAS case requires a 9th-order polynomial. The order of these polynomials ranges from 6 -- 10 for other SARAS beam perturbations. It is important to highlight that these polynomials are tightly constrained to be maximally smooth, permitting only one inflection point. This constraint ensures that these polynomials do not have the flexibility to subsume the spectral features of plausible 21-cm signal profiles even for higher orders, unlike unconstrained polynomials. This is further demonstrated in Section \ref{sec: signal extraction}, where we evaluate 21-cm signal detection prospects with these foreground models.

We find raft height to be the most critical parameter, resulting in relatively high residuals. Notably, in Table \ref{tab: residuals} corresponding to cases 12 and 13, the residual r.m.s values following the modeling are four orders of magnitude higher than the others. 
The observed rise in residuals can be interpreted as follows.
In freshwater, the intrinsic impedance of the medium remains stable due to higher permittivity compared to conductivity , exerting minimal influence on antenna impedance and beam response. However, changes in raft height significantly impact the impedance and beam response as the width of the air-dielectric interface changes. Conversely, in case of tilt-induced variations, since the height variation is uneven, the effects are averaged out.

However, while r.m.s of residuals represent the power contained in the systematics, their spectral profile is of utmost significance. We address this in
Section \ref{sec: signal extraction}, where we jointly model the simulated data with different variations of the global 21-cm signal.

\subsection{Beam derivative as a measure of chromaticity} \label{subsec: chromaticity factor}
\begin{figure}
\centering
\includegraphics[scale=0.50]{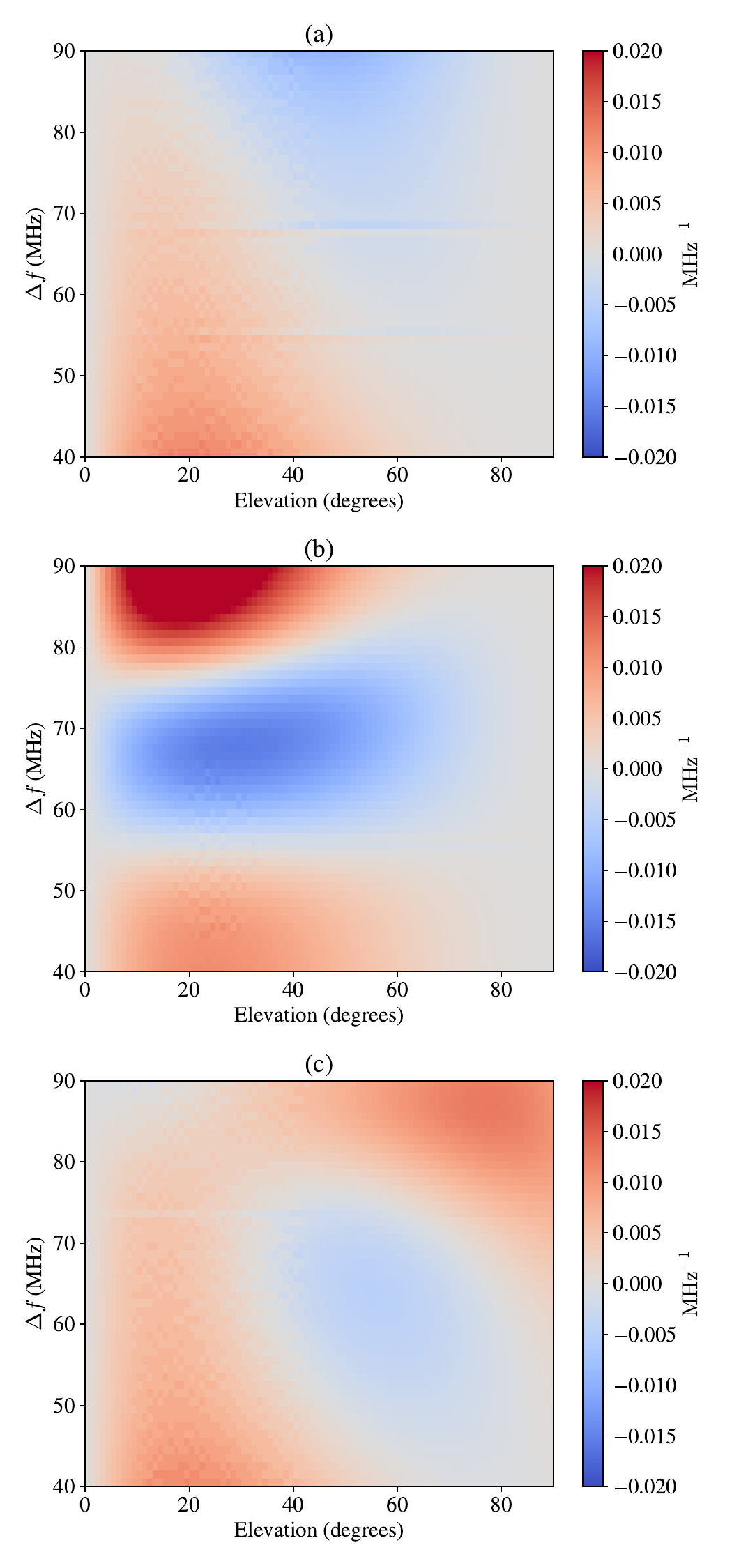}
\caption{Panel (a) shows the beam derivative for the antenna in an ideal environment, panel (b) shows the beam derivative for the case where the raft height reduces to 100 mm, and panel (c) shows the derivatives when the raft gets tilted to 6 degrees from the zenith due to instabilities in the operating environment. The derivatives clearly capture the additional chromaticities introduced due to these perturbations.}
\label{fig: beam derivative plots}
\end{figure}

In Section \ref{subsection : purterbed beam results}, we noted the features arising due to variations in the primary beam. This could be a potential result of beam chromaticity, making a comprehensive analysis of chromaticity essential for a better understanding. While the beam varies with elevation, azimuth, and frequency, it's the latter that can result in the added spectral complexity. We, therefore, define beam derivative along the frequency axis for each elevation and azimuth angle. Such analysis of beam chromaticity has been done in \cite{2016MNRAS.455.3890M}. These beam derivatives are shown in Figure \ref{fig: beam derivative plots} for the baseline case, along with variation in the raft height and tilts. In the subsequent panels, enhanced chromaticity for perturbed cases is obvious from the beam derivative. We quantify the beam derivative into a chromaticity factor. It is computed by estimating zero crossings in the beam derivative at each elevation and azimuth angle. A derivative contrast is computed by finding local maxima and minima between two successive zero crossings, followed by summing it up for all angles as shown in Equation \ref{eq: chromaticity factor}.

\begin{equation}
\label{eq: chromaticity factor}
 \small 
\begin{gathered}
            \textit{Chromaticty} \\[-\jot]
            \textit{factor}
        \end{gathered} =\sum_{\theta = 0}^{\pi/2} \sum_{\phi = 0}^{2\pi}\sum_{i = 0}^{N} \left( \frac{\Delta G_{\substack{local \\ max \ i}}} {\Delta f} - \frac{\Delta G_{\substack{local \\ min \ i}}} {\Delta f} \right ),
\end{equation}

where $N$ is the total number of zero crossings, $\Delta f$ is the frequency resolution, and $\Delta G_{\substack{local \\ max}}$ and $\Delta G_{\substack{local \\ min}}$ are the absolute values of local maxima and local minima of beam derivatives between the zero crossing.

\begin{figure*}
    \centering
    \includegraphics[scale=0.50]{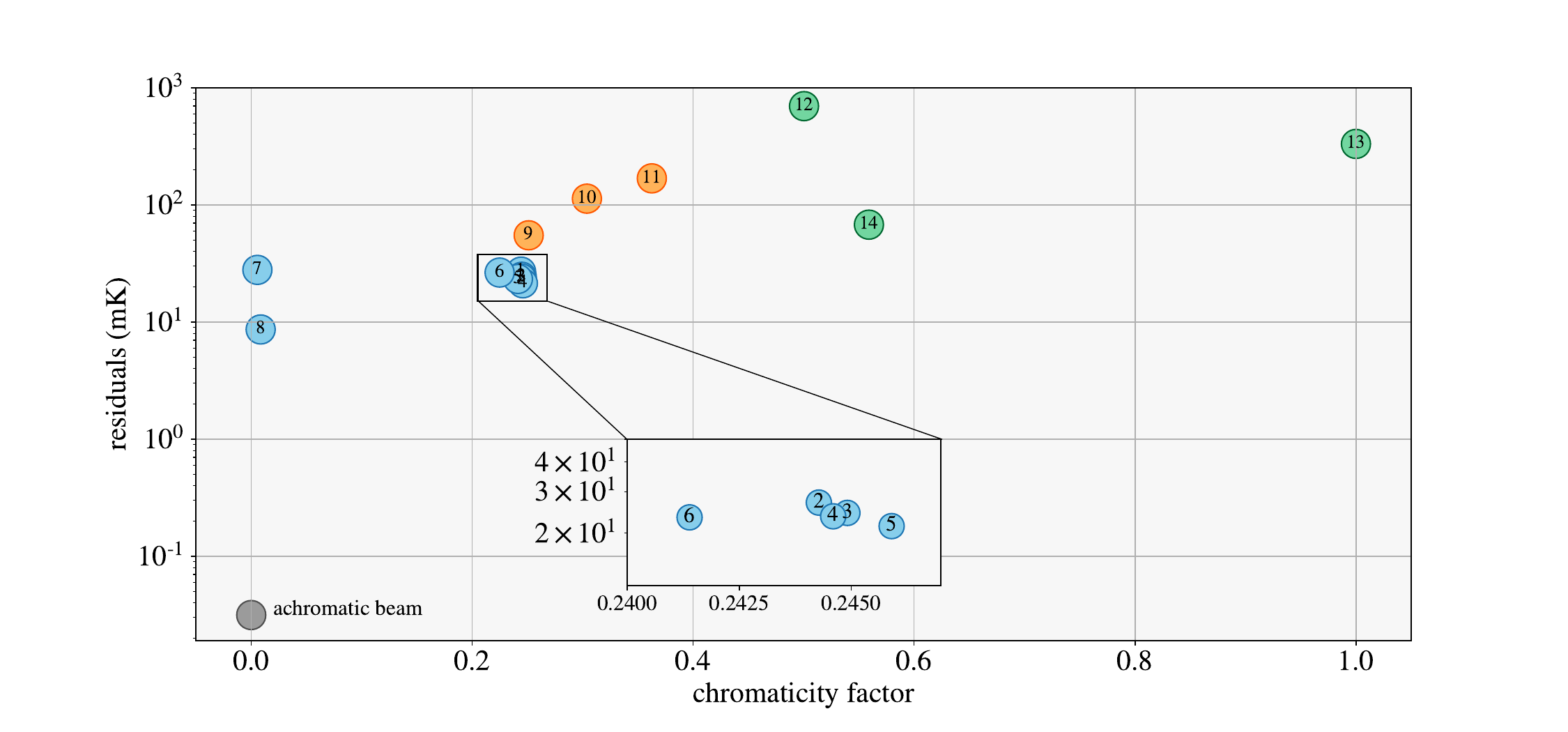}
    \caption{Residual r.m.s. after fitting maximally smooth function with no inflection points versus normalized chromaticity factor for different cases as discussed in Equation~\ref{eq: chromaticity factor}. Numbers show different configurations in Table \ref{tab: residuals}. Conductivity, tilt, and raft height variations are plotted in blue, orange, and green, respectively.}
    \label{fig: chromaticity_fact_vs_res}
\end{figure*}

In Figure \ref{fig: chromaticity_fact_vs_res}, we compare chromaticity factors as computed above to the r.m.s of residuals obtained after fitting out the beam-weighted sky with a maximally smooth function allowing no inflections. Adopting a maximally smooth function with no inflection brings up spectrally rich structures, enabling a true estimate of chromaticity introduced in each case. Note that we normalize the chromaticity factors by the peak chromaticity to allow for relative comparisons. Expectedly, we find correlations between them, further indicating that excess spectral structure arises due to the rate of change of beam in the spectral domain. However, we also notice a non-linear relation between them for raft heights, indicating the importance of forward modeling foregrounds.

\section{Analyzing Variation in Return Loss} \label{sec: S11 correction}
Return loss governs the coupling of power available at the antenna into the receiver chain. We quantify this coupling via reflection efficiency defined as,

\begin{equation}
    \eta_{r} = 1 - |\Gamma_{a}^{2}|
\end{equation}

where $\Gamma_{a}$ is the complex reflection coefficient, which is the linear equivalent of return loss or S11 in the log scale. The observed brightness, $T_{a} = \eta_{r}T_{W}$, clearly shows any spectral features in the return loss are imprinted on the observed data. Although experiments account for this effect, alterations in the antenna configuration can lead to variations in $\eta_{r}$. For example, changes in conditions around the antenna can directly impact $\eta_{r}$. Therefore, any miscalibration due to changed conditions between measurements of $\eta_{r}$ and observations can introduce additional spectral features.

\begin{figure*}
    \centering
    \includegraphics[scale=0.55]{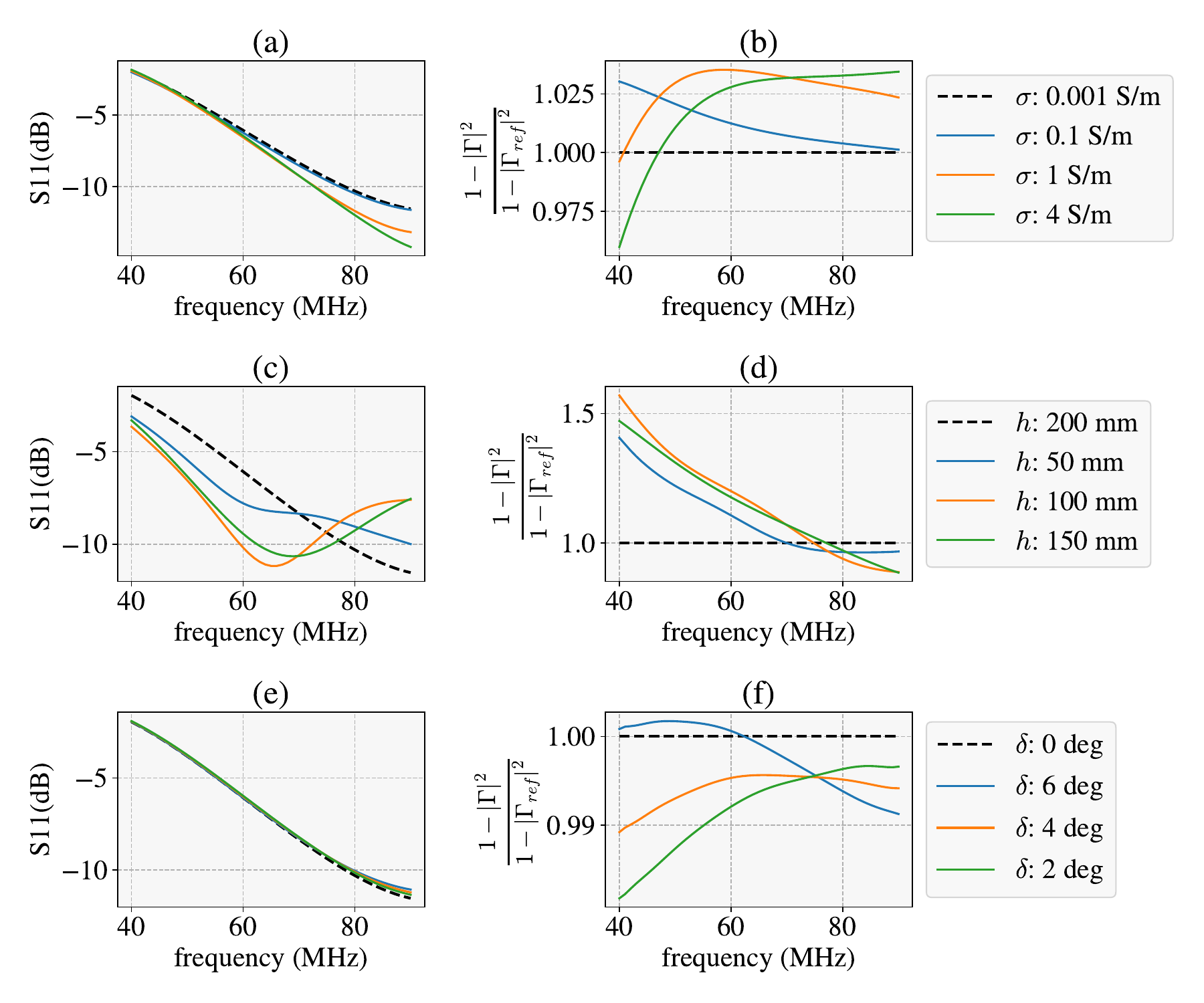}
    \caption{Panels (a), (c), and (e) show the simulated S11 for conductivity, raft height, and raft tilt variations, respectively. Panels (b), (d), and (f) denote the respective S11 miscalibration factors.}
    \label{fig: return_loss_stuff}
\end{figure*}

 This section will assess the systematics introduced by miscalibrated S11 corrections. Simulated plots of S11 for different SARAS cases are shown in Figure \ref{fig: return_loss_stuff}, panel (a), (c), and (e).

\subsection{Data Simulation and Analysis}
We take simulated data for each case just like Section \ref{sec: beam smooth}, but we correct the return loss with the baseline SARAS case, which corresponds to case 1 in Table \ref{tab: residuals}. The simulated dataset is described by the Equation \ref{eq: return_loss_miscaliberated_simulation},

\begin{equation}\label{eq: return_loss_miscaliberated_simulation}
    T_{A}(\nu,t)
    = \left(\frac{1 - |\Gamma(\nu)|^{2}}{1-|\Gamma_{ref}(\nu)|^2} \right) T_{w}(\nu, t), 
\end{equation}

where $T_{w}(\nu, t)$ is taken from Equation 
\ref{eq: simulated_data_cosine}, with $G(\theta, \phi, \nu)$ is being the respective SARAS beam. $\Gamma$ and $\Gamma_{ref}$ are the reflection coefficients from the perturbed case and baseline case, respectively.
Equation \ref{eq: return_loss_miscaliberated_simulation} represents data that takes into account the return losses but is corrected by a mismatched S11 measurement. Such miscalibration is representative of scenarios where there is a temporal variation in $\Gamma$ due to changes in medium property or antenna configuration.

The data is subsequently fitted to the same maximally smooth function as described in Equation \ref{eq: log_log_poly_DC}.

\subsection{Results}

As seen in Table \ref{tab: residuals}, the residuals have clearly increased for each case for a mismatched S11 correction. It must be noted that the residuals we see here are a combined effect of beam chromaticity and mismatched S11 correction.  In certain instances, the residuals spike significantly, reaching as high as 8 K, rendering signal recovery unfeasible. This shows the significance of carrying out in-situ measurements of S11, factoring in the temporal variations that can introduce large miscalibration errors.

\section{extracting global signal} \label{sec: signal extraction}
As seen in Section \ref{sec: beam smooth}, there is an enhancement in spectral complexity of the beam-weighted foregrounds when convolving the radio sky with the SARAS baseline beam and its variations. One inflection point is deemed necessary for modeling the spectrum to reach the noise level. Nevertheless, there are notably elevated residuals for two cases involving raft heights of 150 mm and 100 mm above the water surface.

In the present section, we aim to extend the modeling and investigate the impact of this excess spectral structure on signal detection. We inject a physically plausible 21-cm signal into the data and attempt to recover it jointly with the foregrounds.

\subsection{Data Simulation and Signal Profiles}
Our model for observed sky comprises beam-weighted foregrounds, global 21-cm signal, and thermal noise given by the,
 
\begin{equation}\label{eq: sky_temperature}
T_{B}(\nu,t) = T_{w}(\nu ,t) + T_{21}(\nu) + T_n (\nu),
\end{equation}
where $T_n (\nu)$ is thermal noise. We assume bandpass calibration systematics are sub-dominant.

\begin{figure}
    \includegraphics[scale=0.35]{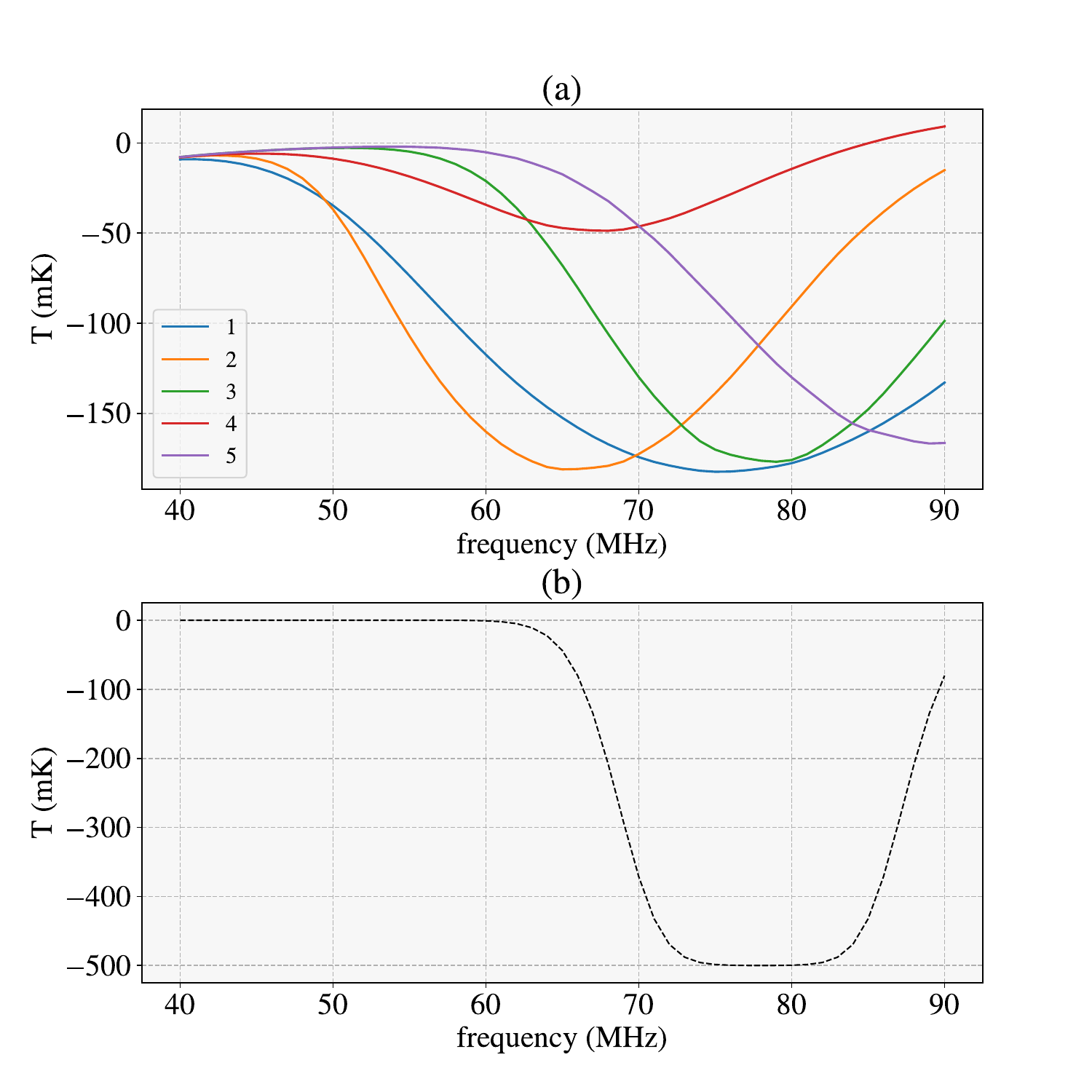}
    \caption{(a) Different 21-cm signals with varying spectral features. Signals 1, 2, and 3 exhibit in-band absorption profiles with deep absorption features. Signal 4, even though in-band, has a shallow absorption feature compared to 1, 2, and 3. Signal 5 exhibits partially out-of-band absorption due to late star formation. Their astrophysical parameters are listed in Table \ref{tab: signal-parameters}. (b) Best-fit flattened profile from \cite{bowman2018absorption}}
    \label{fig: plausible signals}
\end{figure}

It should be noted that the anticipated angular fluctuations in the global signal amount to 1-2 degrees in the sky  \citep{liu2013global}. Conversely, SARAS employs a wide beam characterized by FWHM values spanning from 40 to 47 degrees. In light of this comparison, $T_{21}$ can be generally considered to exhibit negligible angular dependence when contrasted with the SARAS beam width. To simulate the data, we take SARAS beam-weighted sky temperature from GMOSS and add one of the plausible signals. We take a range of theoretically plausible signals with varying spectral features. The selection of signals is governed by spectral profiles, which correspond to different astrophysical parameters such as star formation efficiency $f_{*}$, virial circular velocity $V_{c}$, X-ray efficiency of X-ray sources $f_{X}$, spectral energy density SED, and CMB optical depth $\tau$.  These astrophysical parameters govern their spectral shapes, which were originally discussed in  \cite{2017MNRAS.472.1915C}. We take a subset of signals listed along with their astrophysical parameters in Table \ref{tab: signal-parameters}. The profiles represent a range of detectability over a frequency range of 40-90~MHz, with signals exhibiting absorption features in different parts of the band with varying amplitudes. We have also included the best-fit flattened profile from \citet{bowman2018absorption}. Sharp spectral features represent the strongest detection possibility. Figure \ref{fig: plausible signals} shows different signals taken for this exercise.

\begin{table*}
    \centering
    \begin{tabular}{|c|c|c|c|c|c|}
    \hline
        Signal & $f_{*}$ & $V_c$ (Km s$^{-1}$) & $f_{X}$ & SED & $\tau$\\
        \hline\hline
        1 & 0.05 & 4.2 & 0.1 & 2 & 0.082\\
        \hline
        2 & 0.5 & 16.5 & 0.1 & 2 & 0.082\\
        \hline
        3 & 0.5 & 35.5 & 1 & 1 & 0.066\\
        \hline
        4 & 0.005 & 4.2 & 50 & 1 & 0.066\\
        \hline
        5 & 0.05 & 35.5 & 1 & 1 & 0.083\\
        \hline \hline
    \end{tabular}
    \caption{Astrophysical parameters for the global 21-cm signals shown in Figure~\ref{fig: plausible signals}: star formation efficiency $f_{*}$, virial circular velocity $V_{c}$, X-ray efficiency $f_{X}$, spectral energy density SED, and CMB optical depth $\tau$. The same number identifier will be employed across the paper to denote each specific signal.}
    \label{tab: signal-parameters}
\end{table*}

\subsection{Extraction Methodology and Analysis}
We model the simulated data with a log-log polynomial component  and a scale factor multiplied by the injected global signal as shown in Equation \ref{eq: log_log_poly_scale} 

\begin{equation} \label{eq: log_log_poly_scale}
    T_{ant}(\nu) =  10^{\sum_{n=0}^{m} \alpha_{n} (\log_{10} \nu)^n} + c + k~T_{21cm}(\nu),
\end{equation}

where $T_{ant}(\nu)$ is the antenna temperature, and we use the logarithmic term plus a constant parameter $c$ to model the foregrounds, and $k$ is the scale factor associated with the global signal.

The convergence of the value of $k$ towards unity indicates the successful recovery of the injected signal, while its inconsistency with unity points to either low sensitivity or degeneracy between signal and foreground residuals. Similar method was adopted in \citet{2017ApJ...847...64M}, \cite{2018ApJ...858...54S} and \citet{2022NatAs...6..607S}. 

 We utilize emcee \citep{2013PASP..125..306F} to employ a Bayesian approach to signal detection. Initially, the data is fit to Equation \ref{eq: log_log_poly_scale} using basin-hopping, obtaining the best-fit parameters. These parameter values are subsequently used as initial guesses for MCMC which yield marginalized scale factors with uncertainties along with 2D probability distribution. 
\subsection{Results}

\begin{figure*}[htbp]
    \centering
    \includegraphics[scale=0.30]{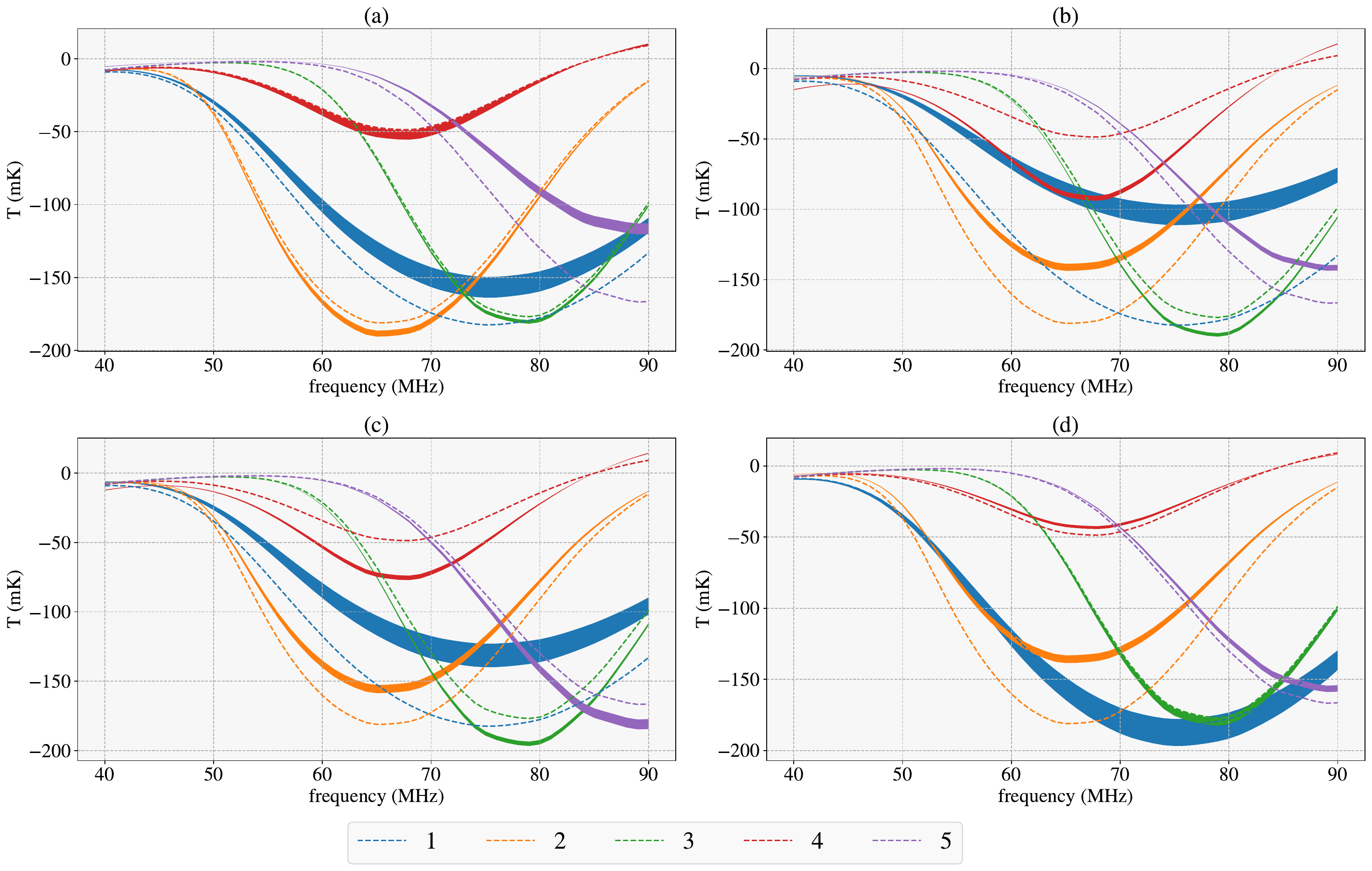}
    \caption{Injected (dashed) and extracted (solid) profiles of different 21-cm signals. Each number represents a unique global signal model as shown in Figure~\ref{fig: plausible signals}. Each panel shows a primary beam specific to a configuration in Table~\ref{tab: residuals}. (a) Baseline SARAS configuration as defined in the first row of Table~\ref{tab: residuals}, (b) higher water conductivity configuration of 0.01 S/m (row 4), (c) case of 6-degree tilt on low conductivity water (row 11), and (d) raft height reduced from 200~mm to 50~mm (row 14).}

    \label{fig: extraced_signals}
\end{figure*}

\begin{table*}
    \centering

    \begin{tabular}{|c|c|c|c|c|c|}
        \hline
        & 1 & 2 & 3 & 4 & 5 \\ \hline
        \multicolumn{6}{|c|}{Conductivity Variation Cases} \\
        \hline
        1 & $0.86^{+0.038}_{-0.040}$ & $1.04^{+0.012}_{-0.011}$ & $1.02^{+0.007}_{-0.007}$ & $1.13^{+0.01}_{-0.01}$ & $0.7^{+0.022}_{-0.030}$ \\ \hline
        2 & $0.79^{+0.042}_{-0.042}$ & $1.07^{+0.007}_{-0.007}$ & $1.05^{+0.006}_{-0.006}$ & $1.52^{+0.018}_{-0.014}$ & $0.69^{+0.018}_{-0.024}$ \\ \hline
        3 & $0.47^{+0.042}_{-0.050}$ & $0.99^{+0.007}_{-0.008}$ & $1.05^{+0.007}_{-0.007}$ & $1.29^{+0.011}_{-0.010}$ & $0.83^{+0.010}_{-0.012}$ \\ \hline
        4 & $0.57^{+0.040}_{-0.040}$ &{$0.78^{+0.014}_{-0.014}$} & $1.07^{+0.007}_{-0.007}$ & $1.89^{+0.041}_{-0.031}$ & $0.85^{+0.012}_{-0.012}$ \\ \hline
        5 & $0.75^{+0.03}_{-0.03}$ & $1.01^{+0.007}_{-0.006}$ & $1.02^{+0.006}_{-0.006}$ & $1.14^{+0.009}_{-0.009}$ & $0.93^{+0.008}_{-0.008}$ \\ \hline
        6 & $0.42^{+0.042}_{-0.050}$ & $1.02^{+0.007}_{-0.008}$ & $1.05^{+0.006}_{-0.006}$ & $1.30^{+0.011}_{-0.011}$ & $0.8^{+0.010}_{-0.011}$ \\ \hline
        7 & $0.74^{+0.037}_{-0.044}$ & $1.08^{+0.007}_{-0.007}$ & $1.07^{+0.006}_{-0.006}$ & $1.39^{+0.014}_{-0.013}$ & $0.72^{+0.012}_{-0.015}$ \\ \hline
        8 & $0.89^{+0.025}_{-0.025}$ & $1.01^{+0.006}_{-0.006}$ & $0.98^{+0.006}_{-0.007}$ & $1.04^{+0.009}_{-0.009}$ & $0.98^{+0.007}_{-0.007}$ \\ \hline
        \multicolumn{6}{|c|}{Tilt Variation Cases} \\
        \hline
        9 & $1.10^{+0.050}_{-0.078}$ & $0.80^{+0.015}_{-0.016}$ & $0.99^{+0.69}_{-0.22}$ & $0.66^{+0.34}_{-0.35}$ & $0.96^{+0.016}_{-0.019}$ \\ \hline
        10 & $1.03^{+0.045}_{-0.047}$ & $1.13^{+0.028}_{-0.028}$ & $0.99^{+0.012}_{-0.012}$& $0.80^{+0.034}_{-0.034}$ & $0.99^{+0.017}_{-0.018}$ \\ \hline
        11 & $0.72^{+0.047}_{-0.046}$ & $0.86^{+0.016}_{-0.017}$ & $1.103^{+0.009}_{-0.009}$ & $1.55^{+0.035}_{-0.032}$ & $1.09^{+0.018}_{-0.025}$ \\ \hline

        \multicolumn{6}{|c|}{Raft Height Variation Cases} \\
        \hline
        12 & $-84.36^{+0.008}_{-0.008}$ & $21.72^{+0.002}_{-0.002}$ & $-0.067^{+0.003}_{-0.003}$ & $5.50^{+0.002}_{-0.002}$ & $-9.18^{+0.002}_{-0.002}$ \\ \hline
        13 & $-174^{+0.005}_{-0.005}$ & $49.06^{+0.002}_{-0.002}$ & $-6.69^{+0.001}_{-0.001}$ & \multicolumn{1}{p{2cm}|}{$-0.318^{+0.008}_{-0.007}$ $-2.336^{+0.009}_{-0.029}$} & $8.06^{+0.092}_{-0.092}$ \\ \hline
        14 & $1.03^{+0.049}_{-0.055}$ & $0.75^{+0.015}_{-0.015}$ & $0.99^{+0.042}_{-0.012}$ & $0.89^{+0.021}_{-0.021}$ & $0.94^{+0.013}_{-0.014}$ \\ \hline

    \end{tabular}
    \caption{Best-fit scale factor after joint fitting beam-weighted foregrounds and 21-cm signals. Each row denotes variation in the primary beam due to change in antenna configuration, and different columns are different models of global 21-cm signal as shown in Figure~\ref{fig: plausible signals}.}
    \label{tab: scale factors}

\end{table*}

The scale factor value recovered in the case of an achromatic beam is consistent with unity for all the injected 21-cm signals, establishing that global 21-cm signal recovery is feasible in the absence of complex spectral features in the foregrounds through joint modeling. 

Scale factors obtained from the joint fitting of different SARAS beam-weighted foreground and global 21-cm signal profiles are reported in Table \ref{tab: scale factors}. In most cases, we find reasonable recovery of the injected signal with a scale factor close to unity. However, some combinations of beams and signals show clear bias. Such bias strongly indicates some degeneracy between foregrounds and global signal profiles. Figure \ref{fig: extraced_signals} shows injected and recovered signal profiles. Upon closer investigation, distinct trends emerge from Table \ref{tab: scale factors}.  Various signal profiles for the same beam have demonstrated varying susceptibilities to spectral distortion, as seen in Figure \ref{fig: extraced_signals}. It is evident from the scale factor values from Table \ref{tab: scale factors} that these biases in recovery depend on the spectral nature of both foregrounds and signal profiles. Signals 1 and 5 consistently exhibit a biased recovery, yielding values below 1 across all conductivity variation cases. The bias ranges from 2~\% to as high as 64~\%. This can be explained by spectrally smooth profiles of these signals, which are most likely to be degenerate with beam-weighted sky profiles. Conversely, signal 4 has shown recovery with positive bias in all conductivity variations. The bias ranges from 4~\% to 89~\% depending on degeneracy. The distortion of profiles for all the signals is quite prominent in cases 12 and 13  of Table \ref{tab: scale factors}, corresponding to reduced raft heights of 100 and 150~mm. These raft height variation cases have proven to generate the most spectral complexities based on their residual values in Table \ref{tab: residuals}. Similarly, antenna tilt cases varying between 2 - 6~degrees exhibit varying levels of distortion, largely influenced by their chromaticity and the spectral profile of the signal itself.
Expectedly, the best-case scenarios emerge for signals 2 and 3, which have distinct absorption features within the band. These signals exhibit the least bias of a few percent for various conductivity cases.

Our further investigation involved a joint fitting absorption profile claimed by the EDGES collaboration \citep{bowman2018absorption}. It's important to highlight that, except for situations where the raft height was set at 100~mm and 150~mm, the injected profile \cite{bowman2018absorption} remained consistently close to unity across all cases. In most instances, its distinctive flattened absorption feature did not exhibit degeneracy with the foreground.

\section{Discussions and conclusions}\label{sec: conclusion}

The paper addressed diverse aspects of systematics that could emerge due to direction-dependent primary beam response and their impact on global 21-cm signal recovery. Our investigation began with understanding the spectral nature of the intrinsic foregrounds and the effect of realistic beam-weighting. This revealed the following key takeaways:

\begin{itemize}
    \item Foregrounds exhibit spectral smoothness to sub-mK level, limited by thermal noise, with a wide achromatic beam. This was demonstrated via modeling foregrounds by maximally smooth functions.
    
    \item However, with several realistic variations of the SARAS beam, the spectral complexity increased to the point that one inflection point was deemed necessary. These spectral features were more prominent in some cases of beam variations, owing to the varying chromatic nature of different beams.

    \item The additional systematics due to the frequency dependence of the primary beam was also correlated with beam derivatives for most variations.

    \item The magnitude of systematics was further exacerbated due to a few percent erroneous correction of S11 or transfer function, highlighting the importance of in-situ measurements.

    \item Ultimately, examining the influence of these additional features on the retrieval of the global 21-cm signals with varying spectral profiles showcased successful recovery of the signals in most cases. In a few cases, biased or distorted extraction indicated some degeneracy between foregrounds, signal profiles, and primary beams. The susceptibility to distortion showed trends based on signal profiles and the spectral complexity of the beam-weighted foregrounds.  We observed that signals with sharp in-band features showed the most robust recovery. Conversely, signals with broad-band or partially out-of-band profiles seemed more prone to bias due to degeneracy with foregrounds and beams.
    
\end{itemize}

The findings discussed above underscore the complex interplay between foregrounds, direction-dependent primary beam, and spectral behavior of 21-cm signals. The profile of 21-cm signal is crucial to consider since a plethora of signal models are still unconstrained. Detectability prospects can significantly vary based on the selection of frequency range and signal model. For example, signals with absorption features, albeit with a lower amplitude, might have higher detection prospects compared to deeper signals with wider widths. Parameter variations with respect to the SARAS radiometer also inform about the allowed degrees of freedom that can lead to robust signal detection.

The results also emphasize the importance of the nature of primary beam chromaticity towards signal detection. Different experiments across the globe employ unique antenna designs with different levels of beam chromaticities. While acknowledging that corrections for primary beams may be non-trivial, the impact of the choice of primary beams on signal detection prospects cannot be overemphasized. Given the relation between the direction-dependent gain of the antenna and its surroundings, it becomes increasingly important to electromagnetically model the response with rigorous characterization of the terrain chosen for conducting experiments.

\section*{Acknowledgments}
The authors sincerely thank the Raman Research Institute and the CMB Distortion Lab members for providing the resources and conducive research environment necessary for this study. The authors are grateful to Mayuri S Rao for her invaluable insights and collaborative efforts throughout this research endeavor. They also thank Vishakha Pandharpure for the cross-verification of several EM simulations used in this work. YA extended special appreciation to Gunjan Tomar, Rajorshi Chandra, Kinjal Roy, and others for their unwavering support during the project.

\vspace{5mm}

\newpage

\bibliography{bibfile}

\end{document}